# Peculiar atomic bond nature in platinum monatomic chains

*Jiaqi Zhang[1], Keisuke Ishizuka[1], Masahiko Tomitori[1], Toyoko Arai[2], Kenta Hongo[3], Ryo Maezono[4], Erio Tosatti[5,6,7], Yoshifumi Oshima[1*]*

1. School of Materials Science, Japan Advanced Institute of Science and Technology, Nomi, Ishikawa 923-1292, Japan

2. Institute of Science and Engineering, Kanazawa University, Kanazawa, Ishikawa 920-1192, Japan

3. Research Center for Advanced Computing Infrastructure, Japan Advanced Institute of Science and Technology, Nomi, Ishikawa 923-1292, Japan

4. School of Information Science, Japan Advanced Institute of Science and Technology, Nomi, Ishikawa 923-1292, Japan

5. International School for Advanced Studies (SISSA), Via Bonomea 265, 34136 Trieste, Italy

6. CNR-IOM Democritos National Simulation Center, Via Bonomea 265, 34136 Trieste, Italy

7. The Abdus Salam International Centre for Theoretical Physics (ICTP), Strada Costiera 11, 34151 Trieste, Italy.

ABSTRACT

Metal atomic chains have been reported to change their electronic or magnetic properties by slight mechanical stimulus. However, the mechanical response has been veiled because of lack of information on the bond nature. Here, we clarify the bond nature in platinum (Pt) monatomic chains by our developed in-situ transmission electron microscope method. The stiffness is measured with sub N/m precision by quartz length-extension resonator. The bond stiffnesses at the middle of the chain and at the connecting to the base are estimated to be 25 and 23 N/m, respectively, which are higher than the bulk counterpart. Interestingly, the bond length of 0.25 nm is found to be elastically stretched to 0.31 nm, corresponding to 24% in strain. Such peculiar bond nature could be explained by a novel concept of "string tension". This study is a milestone that will significantly change the way we think about atomic bonds in one-dimensional substance.

TEXT

**Introduction**

Monatomic chains exhibit unique physical and chemical properties that are sensitive to atomic configurations.[1–5] 5d transition metals such as Ir, Pt, and Au have been found to form stable monatomic chains as a result of relativistic effects.[6–8] These chains show parity oscillation in association with conductance[9–11] as well as quantum switching between 1 and 2 $G_0$ (= $2e^2/h$, where $e$ is the elementary charge and $h$ is the Planck constant) during slight mechanical displacement[12] or electromigration.[13,14] Furthermore, quantized thermal transport in Au atom chains[15,16] and emerging magnetic order in Pt monoatomic chains[17,18] have been reported. The observed behaviors of atomic chains are primarily attributed to muted or strengthened bond in under-coordinated geometries. However, the mechanical properties of the one-dimensional materials reported previously are generally not consistent, and to date there is no consensus on the contributions of the bond strength to the properites.[19]

To better understand the physical and chemical properties of atomic chains it is useful to investigate characteristics such as the bond length and stiffness. Previously, Ibach et al. reported higher vibrational frequency phonons at the edges of vicinal Pt (111) surfaces and suggested that a linear Pt atom chain with high stiffness may be formed at the edge.[20] Some theoretical calculations suggested correlation between the coordination number and force constant.[21] Gilbert et al. examined the physical properties of zinc sulfide nanoparticles while assessing bond characteristics,[22] and Sun proposed that the physical properties of low dimension materials can be universally determined based on the correlation between the coordination number and bond characteristics.[23]

The force constants of atomic bonds have been widely investigated using both Raman and infrared spectroscopy based studies. However, these techniques are not appropriate for investigating nanoscale materials and are unable to determine the properties of individual bonds. Previously, a specially designed holder for transmission electron microscopy (TEM) studies—which was equipped with a silicon cantilever[24]—was used to find the Young's modulus of an Au atom chain. The Young's modulus values measured were in the range 47–116 GPa.[25] This wide range in values was partially attributed to the inherent error in measuring the displacement. Based on trials using a mechanically controllable breaking junction (MCBJ) equipped with a tuning fork, Shiota et al. reported that the stiffness of Pt atom chains depends on the number of atoms in the chain.[19] That is actually compatible with the proposal by Picaud et al.[21] that nanocontact chains are inhomogeneous, thus hard in the middle of chains and soft at the junctions. However, such bond characteristics in the chains were generally not examined. The aim of the current study is to characterize the single bond in Pt monatomic chains using a TEM holder equipped with a quartz length-extension resonator (LER).[26–28] Our developed method also enables us to obtain information on the energy dissipation due to friction or plastic deformation, which is another advantage.

**Results**

The Pt nanocontacts were created by establishing contact with two Pt wires between a counter electrode and the edge of the LER prong (Fig. 1a). Subsequent stretching of the contacts until almost breaking point resulted in Pt monatomic chains that were maintained in several seconds. These chains were free of any contamination because of the ultra-high vacuum conditions in the TEM column. TEM images of the chains were obtained concurrently to the electrical conductance and stiffness measurements. Stiffness values were acquired by the longitudinal oscillation of the LER prong at an amplitude of 27 pm and a resonant frequency of approximately 1 MHz. This small amplitude helped to prevent plastic deformation of the Pt nanocontacts thereby reduced distortion while recording the TEM images. Detail of the experiment method is available in section1 of supporting information. It should be noted that the measured stiffness values contained contributions from the Pt monatomic chains as well as from both bases supporting the chains. Therefore, the contribution from the bases must be removed to obtain the true stiffness of each ideal chains.

Figs. 1b and 1c present the typical time evolution of the conductance and stiffness of a Pt nanocontact, respectively; while Fig. 1d shows a series of TEM images acquired during stretching, which was controlled by the tube piezo at a rate of 0.08 nm/sec. The conductance showed distinct plateaus and stepwise changes over time, corresponding to the elastic and plastic deformation of the Pt nanocontact, respectively (Fig. 1b).[29] Similarly, the stiffness exhibited stepwise change that was associated with the initiation of plastic deformation (Fig. 1c). However, the stiffness values were initially observed to fall rapidly, and then subsequently recovered to a constant value. Such trends have frequently been observed during the stretching of Pt nanocontacts (Figs. 2m-p, video S1). This behavior indicates that the nanocontact fails to exhibit a stable structure immediately upon the initiation of plastic deformation, but that a very short time interval is required to enable the transition to a stable structure via atomic rearrangement. Although such large drops are less visible than the previous results[5,30] because of room temperature, they were in good agreement with long-standing theoretical predictions.[21] The conductance plateau in region C of Fig. 1b at approximately 1.8 $G_0$ indicates the formation of a single Pt monatomic chain.[30] This conductance value was found to be constant over repeated trials, suggesting that the fabrication of Pt monatomic chains is reproducible. Care was taken to ensure that the bases retained a reasonably reproducible shape during the contacting and stretching processes. In this study, approximately 150 Pt monatomic chains were fabricated with two different sets of bases. In the case of one set, the [110] directions of both bases were maintained parallel to the axis of each chain (referred to as [110]–[110] chains hereafter, Figs. 2 a-d). For the other set, the [111] direction of the left-hand base and the [110] direction of the right-hand base were maintained parallel to the axis of the chain (referred to as [111]–[110] chains, Fig. S1, S2, and video S1). In both sets, the Pt monatomic chains were carefully fabricated between the apexes of both bases.

Fig. 2 presents the results obtained from the [110]–[110] chains. The number of Pt atoms comprising the chain—including the two edge atoms on either side—showed variation during the different stretching trials. Hence the chains could be categorized as containing 2–5 atoms. The number of Pt atoms shown in each TEM image (Figs. 2a–d) was identified from the intensity profile along the chain axis (Figs. 2e–h) and the atomic distance in each chain was estimated to be ~0.25 nm. In addition, the experimentally obtained TEM images (Figs. 2a–d) show good agreement with their simulated counterparts, indicating that the number of atoms in each chain was consistent and accurate. Figs. 2i–l show the variation in conductance with time and indicates that the monatomic chains two, three, and four atoms long had a conductance of approximately 1.8 $G_0$, while the five atom chain exhibited a conductance of 1.3 $G_0$. In contrast to previous reports, parity oscillation was not evident in the current study.[9,31] This lack of oscillation is attributed to thermal excitation, since we worked at room temperature. Notably, the conductance values remained unchanged when the chain length was increased by 0.06, 0.08, 0.1, and 0.25 nm for the chains two, three, four, and five atoms long, respectively. Since Pt monatomic chains have both s-state and d-state ($d_{xz}$ and $d_{yz}$) conductance channels, the observed behavior suggests that the d-state contribution to the conductance is very small, as required both theoretically and experimentally to explain the lack of a magnetic signature in Fano shot noise distribution.[32]

The stiffness evolution as a function of time—plotted in Figs. 2m–p—indicates that the stiffness of each chain is affected by the number of atoms. The measured stiffness values were around 18, 11, 8, and 5 N/m for [110]–[110] chains consisting of two, three, four, and five atoms, respectively. The stiffness of each chain was found to show plateau-like behavior during the stretching process, although it sometimes showed slight variations during stretching. The stiffness variations (increases or decreases) may depend on the relationship between the direction of stretching and that of the LER oscillations, as the stiffness reaches maxima when these directions match. However, such variations are not frequently observed in the experiment and the amount of variation is quite small (more data are available in Fig. S3). Therefore, such influences are not considered in the following statistical analysis conducted on the stiffness data.

The stiffness data for the [110]–[110] and [111]–[110] chains is represented as histograms for all of the Pt monoatomic chains in Figs. 3a and 3b. In each histogram, the four peaks color in red, blue, green, and yellow correspond to the monatomic chains consisting of two, three, four, and five Pt atoms. The average stiffness of the [110]–[110] and [111]–[110] chains are shown in Table I. The standard deviations associated with these stiffness data are primarily due to measurement errors such as the signal noise and/or the mechanical instability of the suspended Pt monatomic chains. The stiffnesses of the [111]–[110] chains were slightly lower than those of the [110]–[110] chains with the same number of constituent atoms; that is attributed to differences in the stiffness of their respective bases.

The stiffness of each chain was estimated by assuming that the stiffness of the two bases supporting the chain ($k_b$) was constant, because no change was observed in the configuration of the apex of the pyramidal base with the {111} facets. By considering the dependence of stiffness on coordination number, the assumption that the bond stiffness between the edge atom of the base and the chain atom ($k_{edge}$), is different to that between two neighboring atoms —particularly in the 4 atom ($k_{chain1}$) and 5 atom ($k_{chain2}$) chains—is made, as shown in Fig. 4a.[33] The experimentally measured stiffness values for various Pt monatomic chains ($k_{total(i)}$), where $i$ represents the number of constituent atoms in the chain), can be expressed:

$$\frac{1}{k_{total(2)}} = \frac{1}{k_b} + \frac{1}{k_{edge}} \quad (1)$$

$$\frac{1}{k_{total(3)}} = \frac{1}{k_b} + \frac{2}{k_{edge}} \quad (2)$$

$$\frac{1}{k_{total(4)}} = \frac{1}{k_b} + \frac{2}{k_{edge}} + \frac{1}{k_{chain1}} \quad (3)$$

and

$$\frac{1}{k_{total(5)}} = \frac{1}{k_b} + \frac{2}{k_{edge}} + \frac{2}{k_{chain2}} \quad (4)$$

By calculating the difference between these four values, the base stiffness values, $k_b$, were estimated to be 98.2 N/m for [110]–[110] chains and 65.1 N/m for [111]–[110] chains. Table I shows the stiffness of monatomic chains with 2–5 atoms after removing the stiffness of the base. It is evident that both chain types show similar stiffness values, confirming the accuracy in the estimation of the stiffness of the base. To further confirm this, the base stiffness was computed using another method in which a cylindrical Pt nano-contact was assumed to be suspended between cone-shaped bases on both sides (section 3 in Supporting information). The stiffness of the base in the [110]–[110] chains was estimated to be 95.6±5.8 N/m, while that for the [111]–[110] chains was found to be 55±3.7 N/m. These values show reasonable agreement with those obtained using equations (1)–(4).

The calculated $k_{edge}$, $k_{chain1}$, and $k_{chain2}$ were shown in Table I for [110]–[110] chains and [111]–[110] chains, respectively. It is clear that the variation between $k_{chain1}$ and $k_{chain2}$ for [110]–[110] and [111]–[110] chains is minimal, indicating that the bond stiffness between two neighboring atoms at the middle of the chain is almost the same, regardless of the number of atoms in the chain. Therefore, the bond stiffness between two neighboring Pt atoms at the middle of these chains is approximately 25 N/m. The precision of the stiffness measurement is in the order of 1 N/m for the current experimental conditions[28]. Based on a simple calculation, $k_{chain}$ (~25 N/m) is noted to be slightly higher than $k_{edge}$ (~23 N/m), which is in turn higher than the bulk value (~20 N/m, section 4 in Supporting information). The obtained results agree with previous studies on 5d transition metals such as Ir, Pt, and Au, which suggested that the bond stiffness in low coordination structures (such as chains) is relatively high compared with those in high coordination structures[7,34].

As mentioned above, the Pt monatomic chains show relatively long plateau in time evolution of the conductance and stiffness (Figs. 2i-2p, Fig S3). Since the Pt monatomic chain has stretched at a rate of 0.08 nm/sec, the time interval of the plateau corresponds to the elongation of the chain until breaking. For the sake of simplicity, assuming that initially Pt chain is in the most stable state, the maximum elongation has been statistically investigated for each of the constitute atoms as shown in the histograms of Fig. S11. The maximum elongations have been obtained to be approximately 0.07, 0.13, 0.2, and 0.25 nm for the chains with two, three, four, and five atoms.

Taking into account that the base supporting the Pt chain has also stretched by the elongation, the maximum elongations of the bondedge and the bondchain are almost the same which is estimated to be about 0.06 nm (detail in section 6 of supporting information). Since the equilibrium bond length is 0.25 nm, the maximum elastic strain is approximately 24% for an individual Pt-Pt bond in the chains, which is an extremely large value by considering that the maximum elastic strain in the bulk crystal is less than 5%.

**Discussion**

Quantitative measurement of the bond stiffness is quite important to understand the physical and chemical properties of one-dimensional monatomic chains in states with higher free energy as compared with the bulk configuration. For its existence, a monatomic chain requires two supporting bulk-like electrodes, from where atoms are extracted during mechanical stretching[21]. Remarkably, the free exchange of atoms between bases and chain give rise to an intrinsic string tension[35], which is always a positive quantity, even in the absence of mechanical stretching. This is because an atom always has lower chemical potential in the bulk than that in the chain. The string tension $f$ is the positive work necessary to pull the wire out of the bulk tips per unit length, and is given by[35], $f = (F - \mu N)/L$, where $F$ (a negative quantity) is the free energy of the wire, $N$ is the number of atoms, $\mu$ is the cohesive energy (chemical potential, also negative) of an atom in bulk, and $L$ is the length of the wire. Theoretical research has showed that a tip-suspended wire has a stable configuration at the minimum string tension rather than at that of the free energy $F$ [35,36]. Once stretched, moreover, such a suspended wire can only retain long-lasting metastability so long as the string tension has a positive curvature as a function of increasing length, otherwise it will necessarily break under stretching.[36]

The mechanical properties of monatomic chains have been investigated theoretically by assuming an infinite Pt monatomic chain, as had been done repeatedly in the past.[37] Excellent agreement between the concepts of thermodynamics developed using infinite wires[35] and the real, short experimental ones gives us confidence that this approximation may be carried over even to the short wires considered here. We have obtained the binding energy per Pt atom as a function of the interatomic distance, which can be regarded as the free energy of the chain at 0 K, based on density functional theory (DFT) as shown in Fig. 4b (section 5 in Supporting information). Accordingly, the string tension has also been plotted (Fig. 4b), assuming that µ is the value for bulk Pt (–5.82 eV). The Pt monatomic chain reaches a theoretically stable structure when the interatomic distance is ~0.25 nm corresponding to the local minimum of the string tension, about 4% larger than the total energy minimum. The corresponding equilibrium string tension is about 1.38 eV/Å (2.2 nN), about 40% stronger than the corresponding Au wire.[36] The bond stiffness corresponding to the second derivative of the energy curve at the string tension minimum in Fig. 4b was calculated to be 26.2 N/m. Under stretching, the theoretical bond length and string tension barrier at the breaking point is 0.32 nm and about 2.3 nN, respectively. The bond length at the breaking points in the Pt monatomic chains is estimated to be 0.31±0.02 nm in our experiment. Although the actual string tension cannot be obtained in our system, the excellent agreement of the equilibrium bond length, stiffness and maximum bond length (breaking point) with the experimental results strongly support the effect of string tension of a Pt–Pt bond. The present results, combining precise measurements, theoretical calculations, and careful modeling of nano-contact inhomogeneity, represent a first demonstration of the string tension-governed nanowire mechanics. The additional role of nano-magnetism of Pt monatomic chains proposed theoretically[17,37] and confirmed experimentally[18] at cryogenic temperature, could form the subject of future studies. We have succeeded in identifying individual bond stiffnesses by incorporating the force measurement method based on the FM method. Furthermore, since this method can be used to measure mechanical energy dissipation during TEM observation, we believe that the mechanism of friction and plastic deformation could be clarified on an atomic scale. The overall increase in understanding gained through consideration and control of string tension should also be of immense importance in atomic scale mechanics and electronics with monoatomic chains.

**Conclusion**

In this study, *in situ* TEM coupled with LER based force sensor measurements confirmed the formation of one-dimensional Pt monatomic chains—consisting of two, three, four, or five atoms—between Pt bases with equivalent shapes. The stiffness and electrical conductance of the chains were measured concurrently during TEM observation; both exhibited plateau-like behaviors throughout each elastic stretching stage. By eliminating the stiffness of the bases, the bond stiffness between two atoms at the middle of the chain was estimated to be approximately 25 N/m; regardless of the chain length and the crystal orientation of the suspended-tips. Furthermore, the bond length of 0.25 nm was found to be elastically stretched to 0.31 nm, corresponding to 25% in strain. Such peculiar bond nature proves that string tension is a key concept to understand the physical or chemical properties of one-dimensional system. The present study provides basic information on the microscopic understanding of friction and the development of monatomic and molecular devices.

FIGURES

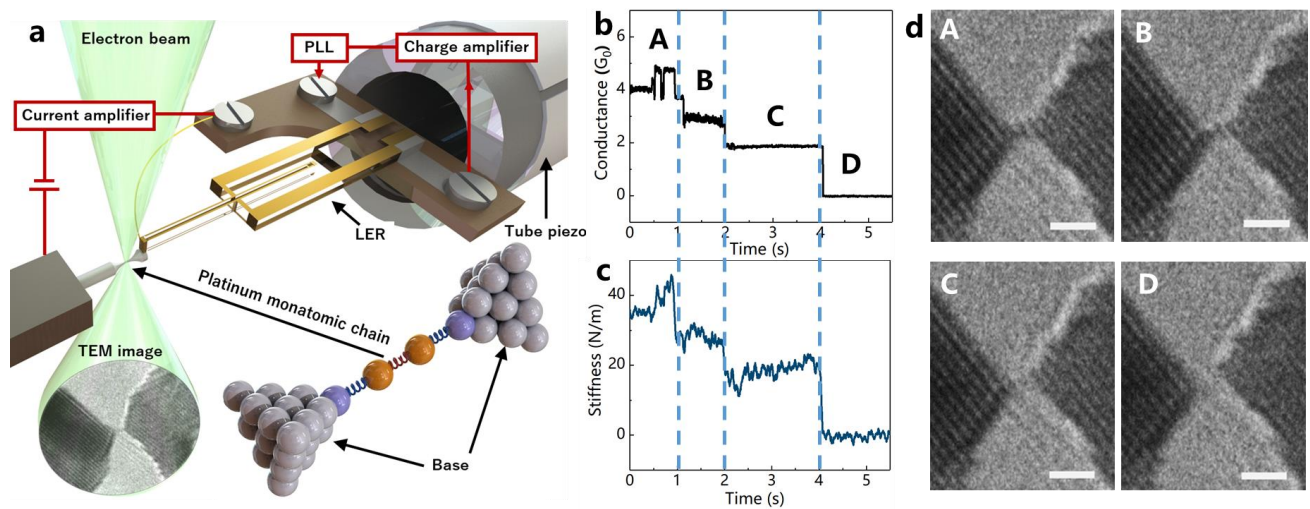

Figure. 1 Schematic illustration of the experimental setup and typical variations of conductance, stiffness, and morphology (as observed by TEM) for Pt monatomic chains during stretching. a, A schematic illustration of the experimental setup. The Pt monoatomic chain was formed between a sharpened Pt wire attached to a quartz length-extension resonator (LER) and another Pt wire fixed to a copper plate. A constant bias, $\Delta V$, was applied between the two Pt wires to determine the conductance of the Pt monoatomic chain. The position of the sharpened Pt wire was controlled in three dimensions by a tube piezoelectric device. The electrode on one side of the LER was excited by applying a sinusoidal voltage, leading to oscillation of the LER. A feedback loop was used to tune the excitation frequency, ω, (via a phase-locked loop; PLL) to ensure that the LER oscillated at its resonant frequency with a constant amplitude[28]. Typical variations over time of the b, conductance and c, stiffness. And d, TEM images captured at moments A, B, C, and D marked in b and c. The scale bars indicate 1 nm.

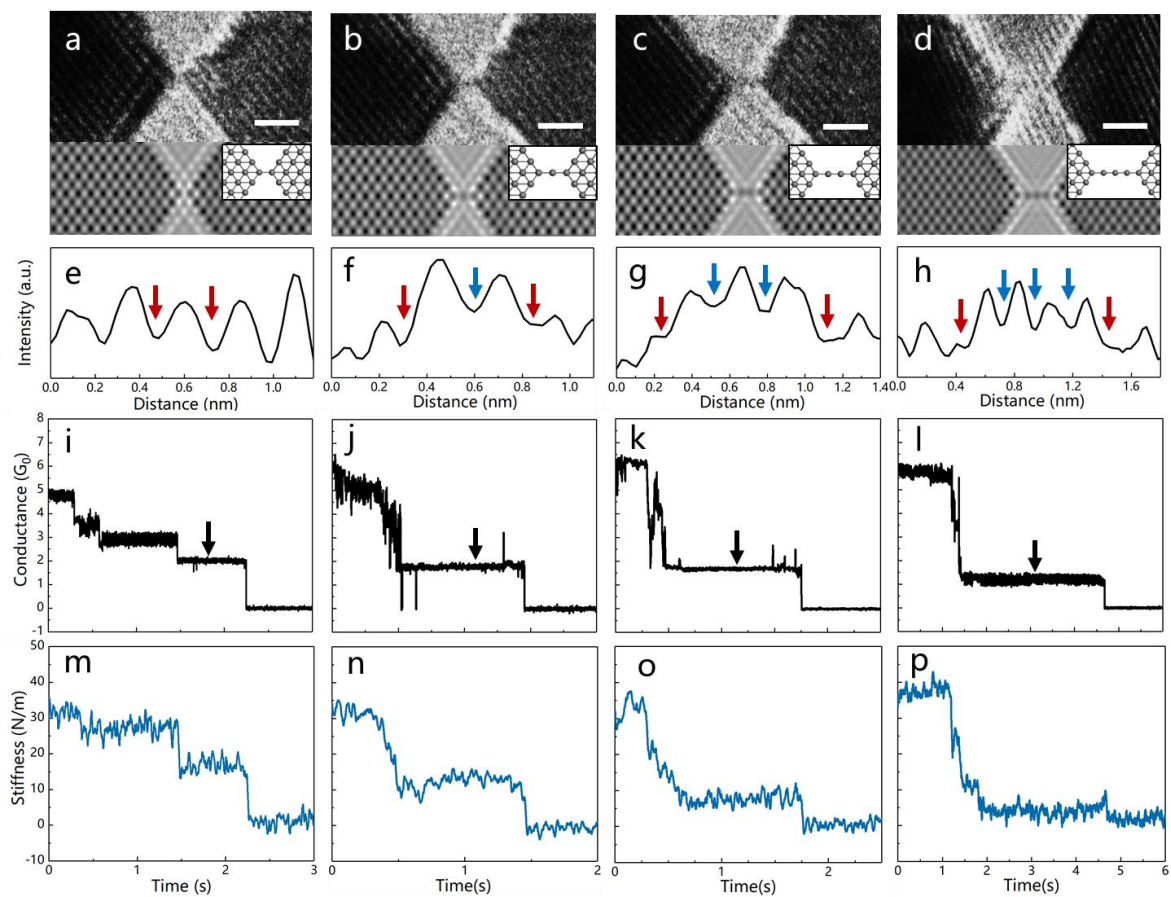

Figure. 2. a–d, TEM images of chains comprising two, three, four, and five atoms, and the respective simulated TEM images of the atomic configurations based on the TEM observations. The scale bars indicate 1 nm. TEM images of the Pt atom contacts were simulated by multi-slice calculations (HREM Research). The atomic positions were derived from the configurations of Pt monatomic chains as obtained from the TEM images. These analyses used an accelerating voltage of 200 kV, a spherical aberration coefficient of 0.7 mm, a chromatic aberration of 1.2 mm, and a defocus value of 50 nm (under-focused condition). e–h, Intensity line profiles along the axes of monatomic chains containing two, three, four, and five atoms. The red and blue arrows indicate the edge and central atoms, respectively. i–l, Variations of the conductance of monatomic chains with two, three, four, and five atoms over time. The arrows indicate the acquisition time of the TEM image. m–p, Variations of the stiffness of monatomic chains with two, three, four, and five atoms over time.

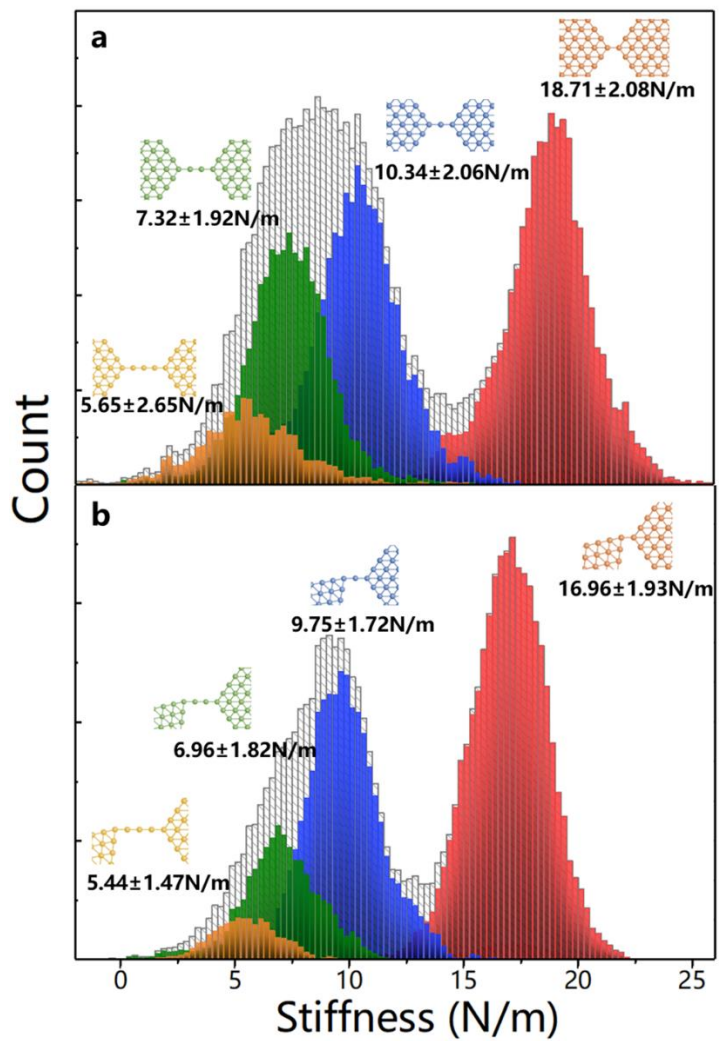

Figure. 3 Stiffness histograms of the Pt monatomic chains. a, Stiffness histograms of the [110]–[110] monatomic chains. b, Stiffness histograms of the [110]–[111] monatomic chains. The gray plot is the total histogram, while the red, blue, green, and yellow plots are the stiffness data for chains of two, three, four, and five atoms, respectively. The structures and average stiffness values are shown beside the corresponding peaks.

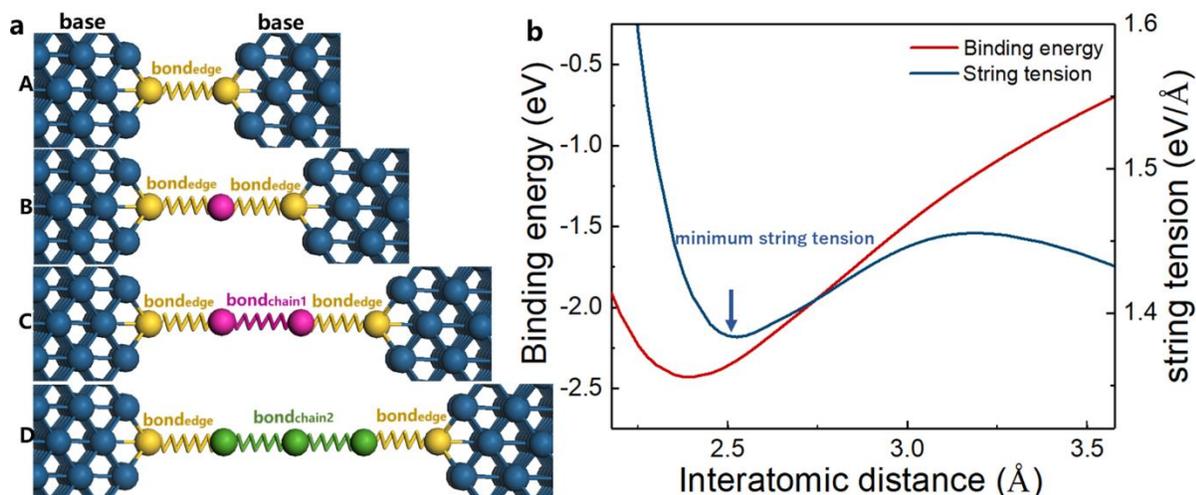

Figure. 4 Models used to calculate bond parameters for the monatomic chains. **a,** Structure models of monatomic chains containing two, three, four, and five atoms, corresponding to one, two, three, and four atomic bonds, respectively. The blue, yellow, pink and green atoms indicate atoms at the base, edge, and two types of chain atom, respectively. The bond of the chain at middle is expressed by $bond_{chain1}$ or $bond_{chain2}$, while the edge bond is expressed by $bond_{edge}$. **b,** Binding energy (red curve) and string tension (blue curve) as a function of the interatomic distance in a monatomic chain with infinite length. The binding energy was obtained by DFT calculation (see details in methods and section 5 in the Supplementary information).

TABLES.

Table I Stiffness of monatomic chains and individual bond

|  | [110]-[110] chains | | [111]-[110] chains | |
| --- | --- | --- | --- | --- |
| Number of atoms in the chain | Measured stiffness ($k_{total}$, N/m) | Stiffness after remove the contribution of base ($k$, N/m) | Measured stiffness ($k_{total}$, N/m) | Stiffness after remove the contribution of base ($k$, N/m) |
| 2 atoms | 18.71±2.08 | 23.1 | 16.96±1.93 | 22.9 |
| 3 atoms | 10.34±2.06 | 11.5 | 9.75±1.72 | 11.5 |
| 4 atoms | 7.32±1.92 | 7.9 | 6.96±1.82 | 7.8 |
| 5 atoms | 5.65±2.65 | 6.0 | 5.44±1.47 | 5.9 |
| Bond type | Bond stiffness (N/m) | | | |
| $bond_{edge}$ | 23.1 | | 22.9 | |
| $bond_{chain1}$ | 25.1 | | 24.3 | |
| $bond_{chain2}$ | 24.9 | | 24.6 | |

The errors in the measured stiffness is the standard deviation of the measured data. In the analysis of individual bonds, we assume that the bond between the edge atom of the base and the chain atom (bond$_{edge}$) is different to that between two neighboring atoms—particularly in the 4 atom (bond$_{chain1}$) and 5 atom (bond$_{chain2}$) chains, as shown in Fig. 4a.

ASSOCIATED CONTENT

Supporting information text(file type, PDF)

Supplementary video (Movie S1) of a [111]-[110] contact (file type, MOV)

AUTHOR INFORMATION

**Corresponding Author**

Yoshifumi Oshima: oshima@jaist.ac.jp

**Author Contributions**

J. Z. performed the experiments. J. Z., K. H., and R. M. conducted the calculations and theoretical modelling. M.T., T.A., E. T. and Y.O. planned the project. All authors discussed the results and prepared the manuscript. The manuscript was written through contributions of all authors. All authors have given approval to the final version of the manuscript.

ACKNOWLEDGMENT

This work was supported by JSPS KAKENHI (grant nos. 18H01825 and 18H03879). The computations in this work were performed using the facilities of the Research Center for Advanced Computing Infrastructure at JAIST. J. Z. thanks the MEXT scholarship and acknowledges financial support by the Sasakawa Scientific Research Grant from The Japan Science Society. E.T. acknowledges support by ERC ULTRADISS Contract No. 834402, and the Italian Ministry of University and Research through PRIN UTFROM N. 20178PZCB5.